\documentclass[aps,prd,twocolumn]{revtex4}
\usepackage{graphicx}
\usepackage{amsmath}
\usepackage{amsfonts}
\usepackage{amssymb}

\newcommand{\TT}{\mathrm{TT}}
\newcommand{\dd}{\mathrm{d}}
\newcommand{\bn}{\mathbf{n}}
\newcommand{\bk}{\mathbf{k}}
\newcommand{\bx}{\mathbf{x}}
\newcommand{\bbv}{\mathbf{v}}
\newcommand{\bN}{\mathbf{N}}
\newcommand{\rg}{\mathrm{g}}
\newcommand{\rs}{\mathrm{s}}
\newcommand{\ru}{\mathrm{u}}
\newcommand{\rd}{\mathrm{d}}
\newcommand{\rr}{\mathrm{r}}
\newcommand{\rt}{\mathrm{t}}
\newcommand{\opt}{\mathrm{opt}}
\newcommand{\tropo}{\mathrm{tropo}}
\newcommand{\GW}{\mathrm{GW}}
\newcommand{\BBN}{\mathrm{BBN}}

\newcommand{\etal}{\textit{et al }}
\newcommand{\arxiv}[1]{\textrm{#1}}
\newcommand{\REVIEW}[4]{\textit{#1} \textbf{#2} {#4} ({#3})}

\newcommand{\BOOKed}[5]{\textit{#1}, #2 ({#3}, {#4}, {#5})}
\newcommand{\Name}[2]{\textrm{#1}~#2}

\begin{document}

\title{Bounds on gravitational wave backgrounds from large distance clock comparisons}

\author{S. Reynaud }
\email{serge.reynaud@spectro.jussieu.fr}
\author{B. Lamine}
\affiliation{Laboratoire Kastler Brossel, UPMC, ENS, CNRS, F75252 Paris}

\author{L. Duchayne}
\author{P. Wolf}
\email{peter.wolf@obspm.fr}
\affiliation{LNE-SYRTE, Observatoire de Paris, CNRS, UPMC, F75014 Paris}

\author{M.-T. Jaekel}
\email{marc.jaekel@lpt.ens.fr}
\affiliation{Laboratoire de Physique Th\'{e}orique de l'ENS, CNRS, UPMC, F75231 Paris}

\date{\today}

\begin{abstract}
Our spacetime is filled with gravitational wave backgrounds that constitute
a fluctuating environment created by astrophysical and cosmological
sources. Bounds on these backgrounds are obtained from cosmological and
astrophysical data but also by analysis of ranging and Doppler signals from
distant spacecraft. We propose here a new way to set bounds on those backgrounds
by performing clock comparisons between a ground clock and a remote
spacecraft equipped with an ultra-stable clock, rather than only ranging to an
onboard transponder. This technique can then be optimized as a
function of the signal to be measured and the dominant noise sources, leading
to significant improvements on present bounds in a promising frequency
range where different theoretical models are competing. We illustrate our
approach using the SAGAS project which aims to fly an ultra stable optical
clock in the outer solar system.
\end{abstract}


\maketitle

\section{Introduction}

The basic observables used for synchronizing remote clocks or ranging to distant
events are built up on electromagnetic signals exchanged between remote
observers and compared to locally available atomic clocks \cite{PetitWolf05}.
This statement applies for example to planetary radar ranging \cite{Shapiro99},
lunar laser ranging \cite{Dickey94,Williams96}, synchronizing orbiting clocks
with Earth-bound standards \cite{Blanchet01} or tracking and navigating probes
in deep space \cite{Anderson96}.

Electromagnetic links feel the gravitational waves (GW) 
and this is currently the main route toward GW detection. It
follows that GW affect ranging and Doppler tracking observables
\cite{Estabrook75,Hellings92}. This effect has been thoroughly studied in
particular with the Pioneer and Cassini probes
\cite{Anderson84,Bertotti99,Tinto02}, leading to constraints on the GW noise
spectrum in some frequency range \cite{Armstrong06}. These studies constitute
one of the windows on the physics of the stochastic GW backgrounds which
permeate our spatio-temporal environment and have an astrophysical or
cosmological origin \cite{Hils90,Giazotto97,Schutz99,Maggiore00,Grishchuk01}.
Their results have to be compared with bounds obtained through different
observations \cite{Abbott06} (more discussions below).

The aim of the present paper is to show that remote clock synchronization is
also affected by GW and might be used to set new bounds. Timing is less
sensitive than ranging at distances shorter than the GW wavelength, but this is
no longer the case at large distances. Furthermore, the timing procedure
can be arranged in order to get rid of uncertainties on the motion of the
remote clock, which might greatly improve the bounds on GW at low frequencies.
The numbers will be discussed below by taking as an example the SAGAS project
which aims at flying ultrastable optical atomic clocks in the outer solar
system \cite{Wolf07}. These numbers heavily rely on the extremely good accuracy
of modern atomic clocks \cite{Diddams04,Hoyt06,Fortier07}.

In the next section, we introduce and compare the basic observables
associated with ranging and timing. We then discuss their sensitivity to
stochastic GW backgrounds as well as the noise sources involved in their
measurement. We finally deduce the constraints on GW backgrounds which
could be drawn from comparisons between accurate clocks at large distances from
each other in the solar system.

\section{Ranging and timing observables}

\begin{figure}[h]
\includegraphics[width=4 cm]{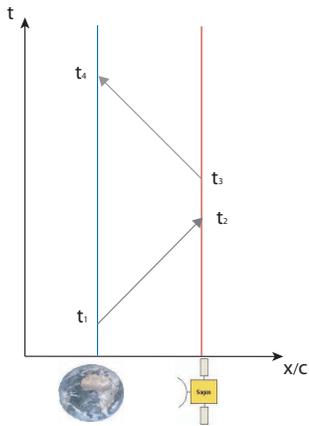}
\caption{Principle of a general two-way link with $t_2\neq t_3$ \label{Figure1} }
\end{figure}

We study the comparison between an atomic clock on board a probe and another 
one colocated with a station on Earth. The clocks are compared using up-
as well as down-links. The uplink signal is emitted from ground at positions 
$(t_1,\bx_1)$ in time and space and received in space at $(t_2,\bx_2)$.
The downlink signal is emitted from space at $(t_3,\bx_3)$ and received on 
ground at $(t_4,\bx_4)$ (see fig. \ref{Figure1}). The up-link is
independent from the down-link i.e. $t_3 - t_2$ can be chosen to take any 
value, positive or negative.

The positions of the emission and reception events are connected by light
cones, which may be calculated through a variety of theoretical methods (see
for example \cite{PetitWolf94_95,LePoncin04,Jaekel06}). These
calculations give solutions which depend on the motions of Earth station and
space probe, as well as on the gravitational field described by the metric. As
the GW are weak modifications of the metric, we treat their effect as a
perturbation of the solutions.

These solutions may be written as relations between clock indications
corresponding to proper times elapsed on ground or space, say $\tau_1^\rg$ and
$\tau_2^\rs$ for the uplink, and $\tau_3^\rs$ and $\tau_4^\rg$ for the downlink
(see fig. \ref{Figure1}). We first define a ranging observable by
\begin{eqnarray}
\tau_\rr &\equiv &-\frac{\tau_3^\rs-\tau_2^\rs}{2}+\frac{\tau_4^\rg -\tau_1^\rg
}{2} \equiv \frac{\tau_\rd +\tau_\ru }{2}  \label{range}
\\
\tau_\ru  &\equiv &\tau_2^\rs -\tau_1^\rg\quad ,\quad \tau_\rd \equiv
\tau_4^\rg -\tau_3^\rs \nonumber
\end{eqnarray}

In the general case up- and downlinks are defined independently from each
other, but a specific configuration of interest is the situation where the
links coincide at the space endpoint ($\tau_2^\rs=\tau_3^\rs \Leftrightarrow
t_2=t_3$). Then $\tau_\rr $ represents a spatial distance corresponding to half
the proper time elapsed on ground during the roundtrip of the signal to and
from the probe. This special case is the only one realized in deep space probes
so far, which were only equipped with a transponder that essentially only
reflects the incoming signal. The range observable (\ref{range}) is unaffected
at lowest order by clock uncertainties in space (when using $t_2=t_3$) and can
even be measured without a good clock on board.

Another observable of great interest is the combination of $\tau_\ru $ and
$\tau_\rd $ defined with the opposite sign
\begin{equation}
\tau_\rt \equiv \frac{\tau_3^\rs +\tau_2^\rs }{2}-\frac{\tau_4^\rg +\tau_1^\rg
}{2} \equiv \frac{-\tau_\rd +\tau_\ru }{2} \label{timing}
\end{equation}
This timing observable (\ref{timing}) is unaffected at lowest order by
uncertainties in the motion of the probe (when $t_2=t_3$). For a probe equipped
with a clock and a two-way system, one can choose to use either (or both) of
the observables (\ref{range}) and (\ref{timing}) with a free choice of the
value of $t_3-t_2$ in order to optimize the measurement depending on the signal
to be measured and the noise affecting the measurements. This is not the case
for probes equipped only with a transponder, which are limited to the special
case of (\ref{range}) with $t_2=t_3$.

We also introduce the time derivatives of (\ref{range}) and (\ref{timing})
\begin{equation}
y_\rr \equiv \overset{\cdot }{\tau}_\rr =\frac{y_\rd + y_\ru }{2} \quad ,\quad
y_\rt \equiv \overset{\cdot }{\tau}_\rt =\frac{-y_\rd +y_\ru }{2}
\label{timederiv}
\end{equation}
The dot symbol represents here a derivation with respect to a commonly defined
time $t$, chosen for any convenient argument. Note that $y_\rr $ is directly
related to the Doppler tracking observable, which has been over the years the
main source of information on the navigation of remote probes
\cite{Asmar05}. Meanwhile $y_\rt $ is directly related to the frequency
comparison of distant clocks, the so-called syntonization observable. The
variations of these quantities can be evaluated in the framework of a
linearized approximation with a reasonably good approximation in the solar
system. A more precise evaluation would be easy by using available methods
\cite{PetitWolf94_95} and it would not change the qualitative
discussions presented below.

The variations in (\ref{timederiv}) are due to the effect of motion of the
probe and ground station and perturbing effects like atmospheric delays, clock
noise, etc... (see sect. 4) on one hand, and to the integrated effect of
gravity along the propagation of the electromagnetic link on the other hand.
The latter is given by an integral along the up- or downlink paths [u] and [d]
\begin{equation}
\delta \tau_{\ru,\rd}=-\frac{1}{2c}\int_{[\ru,\rd]} h_{ij}^\TT \frac{\dd
x_{\ru,\rd}^i}{\dd\sigma }\frac{\dd x_{\ru,\rd}^{j}}{\dd\sigma } \dd\sigma
\label{upordownlink}
\end{equation}
$h_{ij}^\TT$ is the metric perturbation ($h_{\mu\nu}\equiv g_{\mu\nu}-
\eta_{\mu\nu}$ with $g_{\mu\nu}$ and $\eta_{\mu\nu}$ the metric
and Minkowski tensors) in the transverse traceless (TT) gauge;
$\sigma $ is the affine parameter along the path measured as a length,
and $\frac{\dd x}{\dd\sigma }$ the electromagnetic wavevector reduced so that
its time component is unity.

\section{Sensitivity to stochastic GW backgrounds}

We now evaluate the effect of stochastic GW backgrounds as sources of noise on
the electromagnetic links. For simplicity, we consider the background to be
isotropic and unpolarized. We first recover well known results for the ranging
case with $t_2=t_3$ and then discuss the timing observable and the general case
$t_2\neq t_3$.

To this aim, we introduce a plane wave decomposition of the GW background with
wavevector $\bk=k\bn$ ($k=\omega/c$ the modulus of $\bk$ and $\bn$ its
direction)
\begin{equation}
h_{ij}^\TT(t,\bx)=\int\frac{\dd^3\bk}{(2\pi)^3}\,h^\TT_{ij}[k\bn]\,e^{-i\omega
t+i\bk\cdot\bx}+c.c.
\end{equation}
We then write the Fourier component $\delta\tau[\omega]$ 
of the phase shifts appearing in eq.(\ref{upordownlink}).
At each frequency $\omega$, this component is an integral over the direction
$\bn$ of GW plane waves with wavevectors $\bk=(\omega/c)\bn$ weighted by 
sensitivity amplitudes.
Considering as an example propagation along axis 1, we get 
\begin{equation}
\delta \tau[\omega]=-\frac{\omega}{2\pi
c^3}\int\frac{\dd^2\bn}{4\pi}\frac{h^\TT_{11}[k\bn]}{1-\mu^2} \,\beta[k\bn]
\label{fourierq}
\end{equation}
The sensitivity amplitudes $\beta[k\bn]$ depend on the frequency $\omega$ and
the parameter $\mu\equiv \bn_1$, which is the component of $\bn$ along the
direction of propagation of the electromagnetic signal (here the axis 1). For a
signal emitted at $(t_1,\bx_1)$ and received at $(t_2,\bx_2)$ (uplink on figure
1) the sensitivity amplitude is given by~\cite{Jaekel94}
\begin{equation}
\beta[k\bn]=\frac{1+\mu}{-i}\left(e^{-i\omega
t_2}e^{i\bk\cdot\bx_2}-e^{-i\omega t_1}e^{i\bk\cdot\bx_1}\right)
\label{betagen}
\end{equation}

We will consider for simplicity the case of a stationary, unpolarized and
isotropic background. The background may thus be characterized by a spectral
density $S_\GW[\omega]$ giving the strain noise at a space point $\bx$
\cite{Reynaud07}
\begin{equation}
\langle h^\TT_{11}(t,\bx)h^\TT_{11}(0,\bx)\rangle= \frac{4}{3}\int_0^\infty
\frac{\dd\omega}{2\pi}S_\GW[\omega]\cos(\omega t)
\end{equation}
The fluctuations of $\tau$ are finally characterized by a noise spectrum
$S_\tau[\omega]$ such that \cite{Reynaud07}
\begin{eqnarray}
&&\langle \delta \tau(t)\delta \tau(0)\rangle=\int_0^\infty
\frac{\dd\omega}{2\pi} S_\tau[\omega]\cos(\omega t)\nonumber\\
&&S_\tau[\omega]=\frac{5}{8\omega^2}\,b[\omega]S_\GW[\omega]\label{sq}
\end{eqnarray}
The dimensionless function $b$ is obtained by averaging $|\beta[k\bn]|^2$
over the direction $\bn$ of the GW wavevector
\begin{equation}
b[\omega]\equiv\left<|\beta[k\bn]|^2\right>_\bn=\int_{-1}^{+1}
\frac{\dd\mu}{2}\,|\beta[k\bn]|^2\ \label{defb}
\end{equation}

\subsection{Up- and downlinks with $t_2=t_3$}

We first discuss the special case $t_2=t_3$. The sensitivity amplitudes for the
up- and down- links are obtained directly from (\ref{betagen}) by fixing the
origin of coordinates $t_2=t_3=\bx_2=\bx_3=0$ yielding
\begin{eqnarray}
&&\beta _\ru =\frac{1+\mu}{-i} \left(1-e^{i\left( 1-\mu \right) \omega T
}\right)\nonumber\\
&&\beta _\rd =\frac{1-\mu}{i} \left(1-e^{-i\left( 1+\mu
\right) \omega T }\right) \label{defbeta}
\end{eqnarray}
We have introduced shorthand notations for the propagation time $T$ and the
cosine $\mu$ (defined for the uplink)
\begin{equation}
T \equiv t_2-t_1 = t_4-t_3
\end{equation}
Throughout the paper we assume that the relative motion of the space probe and
the Earth during signal propagation is negligible (so that $t_2-t_1 =
t_4-t_3$). More general expressions for the situation where this is not the
case can also be obtained, but are beyond the scope of this work. The up- and
downlink expressions are exchanged by taking opposite signs for the cosine
$\mu$, the propagation time $T$ and the global expression.

The resulting $b-$function is already known \cite{Jaekel94}
\begin{equation}
b\equiv b_\ru =b_\rd = 2\left( \frac{4}{3}+\frac{\sin \left( 2\omega T \right)
-2\omega T }{\left( \omega T \right) ^{3}}\right) \label{valb}
\end{equation}
For obvious symmetry reasons, it has the same form for up- and downlinks.

\begin{figure}[h]
\includegraphics[width=6 cm]{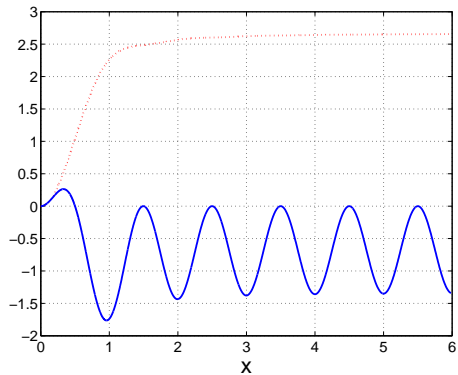}
\caption{$b$ (red-dotted) and $c$ (blue-solid) as functions of $x\equiv
\protect\omega \protect T /\protect\pi $
\label{Figure2} }
\end{figure}

The variation of $b$ versus the dimensionless parameter $x\equiv \omega T /\pi
$ is shown as the red-dotted curve on Figure 2. The blue-solid curve on Figure 2
represents the covariance function $c$ which describes the correlation of
the up- and downlinks ($^\ast$ denotes complex conjugation)
\begin{equation}
c[ \omega ] \equiv \frac{1}{2}\left\langle \beta _\ru [ k\bn ] \beta _\rd ^\ast
[ k\bn ] +\beta _\rd [ k\bn ] \beta _\ru^\ast [ k\bn ] \right\rangle _{\bn}
\label{defc}
\end{equation}
The average over $\bn$ can be evaluated as
\begin{eqnarray}
\label{valc}
&&c [ \omega ] =-4\gamma [ \omega ] \cos(\omega T) \\
&&\gamma [ \omega ] \equiv
 \frac{\cos (\omega T) }{3} + \frac{\cos (\omega T) }{\left( \omega T \right) ^{2}}
-\frac{\sin (\omega T) }{\left( \omega T \right) ^{3}} \nonumber
\end{eqnarray}

We conclude this section by discussing qualitatively the shapes of the two
curves $b$ and $c$. We first notice that $\beta _\ru $ and $\beta _\rd$ tend to
become identical at the limit of short distances or low frequencies ($\beta
_\ru \simeq \beta _\rd \simeq \left(1-\mu^{2}\right) \omega T $ for $\omega T
\ll 1$), so that $b$ and $c$ show the same behaviour $\simeq \frac{8}{15}\omega
^{2}T ^{2}$. The noise spectrum (\ref{sq}) is thus reduced to the simple form
(corresponding to $\delta \tau_\ru \simeq \delta \tau_\rd \simeq
-\frac{T }{2}h_{11}$)
\begin{equation}
S_\tau[ \omega ] =\frac{T ^2}{3}S_{\GW}[ \omega ] \quad ,\quad \omega T \ll 1
\end{equation}
At the high-frequency or large distance limit $\omega T \gg 1$ in contrast, $b$
goes to a constant so that
\begin{equation}
S_\tau[ \omega ] =\frac{5}{3\omega ^{2}}S_{\GW}[ \omega ] \quad ,\quad \omega T
\gg 1
\end{equation}
Meanwhile, the correlation $c$ between up- and downlinks remains sensitive to
the distance even at large distances. This is simply due to the fact that
$\left( \beta _\ru \beta _\rd^{\ast }+\beta _\rd \beta _\ru ^{\ast }\right) /2$
contains a part $\left( \mu^{2} -1\right) \left( 1+\cos \left( 2\omega T
\right) \right) $ which is not blurred by the integration over $\mu$. We also
note that $c$, which is positive at low frequencies ($\omega T \ll 1$), is
negative at high frequencies ($\omega T >\frac{\pi }{2}$).

\subsection{Ranging and timing with $t_2=t_3$}

We repeat now the same discussion in terms of the ranging and timing
observables. It is clear from (\ref{range}) and (\ref{timing}) that one can
write expressions similar to (\ref{fourierq}) for $\delta \tau_\rr $ and
$\delta \tau_\rt$ with the following sensitivity amplitudes
\begin{equation}
\beta _\rr =\frac{\beta _\ru +\beta _\rd }{2}\quad ,\quad \beta _\rt
=\frac{-\beta _\ru +\beta _\rd }{2} \label{betart}
\end{equation}
The noise spectra have the same form as (\ref{sq}) with the sensitivity
functions
\begin{equation}
b_\rr [ \omega ] =\left\langle \left\vert \beta _\rr [ k\bn ] \right\vert
^{2}\right\rangle _{\bn} \quad ,\quad b_\rt [ \omega ] =\left\langle \left\vert
\beta _\rt [ k\bn ] \right\vert ^{2}\right\rangle _{\bn} \label{brt}
\end{equation}
which can be written in terms of the already discussed functions $b$ and $c$
\begin{equation}
b_\rr [ \omega ] =\frac{b[ \omega ] +c[ \omega ] }{2} \quad ,\quad b_\rt [
\omega ] =\frac{b[ \omega ] -c[ \omega ] }{2}
\end{equation}

It has also to be stressed that the correlation between the ranging and timing
variables vanishes, as can be shown through an explicit calculation. As a
matter of fact, the sensitivity amplitudes can be written as
\begin{eqnarray}
\beta _\rr  &=&\left( \sin \omega T -\mu \sin \mu \omega T \right.\nonumber \\
&&\left. -i\mu \left( \cos
\omega T -\cos \mu \omega T \right) \right)
\exp ^{ -i\mu \omega T }  \nonumber \\
\beta _\rt  &=&\left( \sin \mu \omega T -\mu \sin \omega T \right.\nonumber \\
&&\left. +i \left( \cos
\omega T -\cos \mu \omega T \right) \right) \exp ^{ -i\mu \omega T }
\end{eqnarray}
and it turns out that $\left( \beta _\rr \beta _\rt ^\ast +\beta _\rt \beta
_\rr ^\ast \right) /2$ is odd in $\mu$ and vanishes after the angular
integration. Alternatively the fact that the correlation between $\delta
\tau_\rr $ and $\delta \tau_\rt $ vanishes can be directly inferred from the
already discussed property $b_\ru =b_\rd $, which was attributed to a symmetry
between up- and downlinks. It means that $\delta \tau_\rr $ and $\delta
\tau_\rt $ appear as intrinsic and independent stochastic fluctuations of the
positions in space-time of the end-points.

\begin{figure}[h]
\includegraphics[width=6 cm]{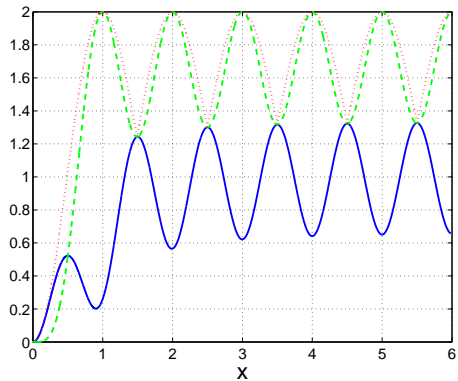}
\caption{$b_\rr $ (blue-solid), $b_\rt $ (green-dashed) and $b_\opt$
(red-dotted, see section 3.3) as functions of $x\equiv \omega T /\pi$
\label{Figure3} }
\end{figure}

The explicit expressions of the functions $b_\rr $ and $b_\rt$ can be obtained
through an explicit calculation or alternatively deduced from (\ref{valb}) and
(\ref{valc})
\begin{eqnarray}
b_\rr [ \omega ] &=& \frac{3-\cos \left( 2\omega T \right) }{3}-\frac{3+\cos
\left( 2\omega T \right) }{\left( \omega T \right)
^{2}}+\frac{2\sin \left( 2\omega T \right) }{\left( \omega T \right) ^{3}}  \notag \\
b_\rt [ \omega ] &=& \frac{5+\cos \left( 2\omega T \right) }{3}-\frac{1-\cos
\left( 2\omega T \right) }{\left( \omega T \right) ^{2}} \label{brtexplicit}
\end{eqnarray}
The associated plots are shown for $b_\rr $ (blue-solid) and $b_\rt $ (green-dashed) on
Figure \ref{Figure3}.

Since $\beta _\ru $ and $\beta _\rd $ tend to become identical at the limit of
short distances or low frequencies, it follows that the GW affect essentially
the ranging observable. We effectively obtain in the limit $\omega T \ll 1$ a
much larger value for $b_\rr \simeq 8\left( \omega T \right) ^{2}/15$ than for
$b_\rt \simeq 2\left( \omega T \right) ^{4}/15$. This is obviously the reason
why this case has been much more studied than the timing case. But is also
clear that this is no longer the case for arbitrary frequencies. In particular
we know that $c$ is negative at frequencies $\omega T >\frac{\pi }{2}$
so that $b_\rt $ is larger than $b_\rr $ in this frequency range. We also
notice that the oscillations in $b_\rr $ and $b_\rt $ persist at large
frequencies or large distances with simple behaviours $b_\rr \simeq 1-\cos
\left( 2\omega T \right) /3$ and $b_\rt \simeq 5/3+\cos \left( 2\omega T
\right) /3$ at $\omega T \gg 1$. While this merely reflects the discussion
already devoted to $c$ it is worth noting that the oscillations tend to
disappear in the sum of $b_\rr $ and $b_\rt $ that is also the single-link
expression $b$.

\subsection{General case $t_2\neq t_3$}

For a spacecraft equipped only with a transponder the only observable that can be
obtained is the ranging $\tau_\rr $ (or its derivative $y_\rr $) for the
special case $t_2= t_3$ (cf Fig. 1). An onboard clock enables one to use also
the timing observable $\tau_\rt $, and additionally provides the possibility to
freely choose the ground and onboard measurements, which are combined to form
$\tau_\rr $ or $\tau_\rt $, ie. to freely choose the value of $t_3-t_2$. We
will show below that this choice can be used to optimize the sensitivity of the
observables to a particular signal (in the present case the GW background) for
a given measurement noise spectrum.

The general expression (\ref{betagen}) for the one-way sensitivity amplitude
can be used directly in (\ref{betart}) and (\ref{brt}), to obtain the
sensitivity functions $b_{\rr,\rt}[\omega]$ for an up- and downlink separated
by $T_{23}\equiv t_3-t_2$
\begin{eqnarray}
&&b_\rr [\omega] = \frac{b}{2}-2\gamma\cos\left(\omega T_{23}+\omega T\right)
\nonumber\\
&&b_\rt [\omega] = \frac{b}{2}+2\gamma\cos\left(\omega T_{23}+\omega T\right)
\label{brtgeneral}
\end{eqnarray}
with $b$ given in (\ref{valb}) and $\gamma$ in (\ref{valc}).
The special case $t_2= t_3$ (equations
(\ref{brtexplicit})) is recovered when setting $T_{23}=0$ in
eq.(\ref{brtgeneral}). These general
sensitivity functions can now be used in the data analysis to choose the
optimum value of $T_{23}$ for each frequency $\omega$ as a function of the link
noise and signal travel time $T$. Figure 3 shows the optimized sensitivity
function $b_\opt$, calculated by maximising either of the equations
(\ref{brtgeneral}). One clearly obtains $b_\opt\geq b_\rr,b_\rt$.

\section{Measurement noise}

In the following, we take the SAGAS (Search for Anomalous Gravity using
Atomic Sensors) project \cite{Wolf07} as an example to illustrate the
advantages and versatility provided by missions with an onboard clock
and independent up- and downlinks. The SAGAS project proposes to fly a highly
stable and accurate optical atomic clock and atomic accelerometer on an escape
orbit in the solar system, up to a distance of 50 AU and beyond. It will
use a continuous optical link for clock comparison, navigation, and data
transfer, together with an X-band radio link as a backup. Science objectives
are centered on tests of fundamental physics, in particular gravity on solar
system scales and the exploration of the outer solar system, in particular the
Kuiper belt.

The optical link uses continuous transmission of a laser at $f_0 \approx 444$~THz 
in both directions (up and down) with 1~W at emission, a 40~cm
telescope on board the satellite and 1.5~m telescopes on the ground. 
Numerous perturbations on the link (atmospheric and instrumental losses,
received photon flux and shot noise, stray light, etc...) are discussed in 
\cite{Wolf07}, section 3.3.4. The fundamental science measurements of
SAGAS are the frequency difference between a local laser (optical clock) and 
an incoming laser beam at the same nominal frequency $f_0$, both on
board (up-link) and on the ground (down-link), sampled at 0.01 Hz (see fig. \ref{Figure1}). 
The measurements thus correspond to the observables
$y_\ru $ and $y_\rd $ defined in (\ref{timederiv}). Including only terms whose noise 
contribution plays a significant role they can be expressed as
\begin{eqnarray}
  y_\ru  &=& \frac{f_\rs (t_2)-f_\rg (t_1)}{f_0}
  + \bN_\ru \cdot\frac{\bbv_\rs (t_2) - \bbv_\rg (t_1)}{c}\nonumber \\
  &&+\Delta y_\tropo(t_1) \nonumber \\
  y_\rd  &=& \frac{f_\rg (t_4)-f_\rs (t_3)}{f_0}
  + \bN_\rd \cdot\frac{\bbv_\rs (t_3) - \bbv_\rg (t_4)}{c}\nonumber \\
  &&+\Delta y_\tropo(t_4) \label{df/f}
\end{eqnarray}
where $f_{\rs,\rg}$ are the frequencies of the space/ground laser (optical
clock), $\bbv_{\rs,\rg}$ the associated velocity vectors, $\bN_{\ru,\rd}$ the
direction vectors of up- and down-links $\left(\bN_\ru \equiv\frac{\bx_\rs
(t_2)-\bx_\rg (t_1)}{\|\bx_\rs (t_2)- \bx_\rg (t_1)\|}\right.$, $\left.\bN_\rd
\equiv\frac{\bx_\rs (t_3)-\bx_\rg (t_4)}{\|\bx_\rs (t_3)- \bx_\rg
(t_4)\|}\right)$, and $\Delta y_\tropo$ the frequency change of the signal due
to it crossing the Earth's troposphere.

The noise coming from the different terms in (\ref{df/f}) can be described
equivalently by a power spectral density (PSD) $S_y(f)$ or an Allan variance
$\sigma_y(\tau)$. In the following we consider the simple cases of a white
frequency noise (terms proportional to $h_0$) and of a flicker frequency noise
(terms proportional to $h_{-1}$) with the translation rule \cite{Barnes71}
\begin{eqnarray}
\label{PSDtoAllan}
    &&S_y(f)= \left( h_0 f^0 + h_{-1} f^{-1} \right) /{\rm Hz}
\nonumber\\
    &&\sigma_y^2(\tau)= {h_0\over 2\tau} + 2\ln2 h_{-1}
\end{eqnarray}
Here the frequency $f$ is in Hz and the integration time $\tau$ in s
(we have kept the notations $h_0$ and $h_{-1}$ used in time and frequency metrology
and which should not be confused with the metric perturbations $h_{ij}^\TT$). We first
discuss the different noise contributions on a single link and then evaluate
the noise on the combined observables $y_\rr $ and $y_\rt $ defined in
(\ref{timederiv}) for the general case $t_2\neq t_3$.

\subsection{One-way link measurement noise}

The space clock fractional frequency stability, as specified in \cite{Wolf07},
corresponds to a white frequency noise $\sigma_y(\tau)=1\times 10^{-14}
\tau^{-1/2}$ and an accuracy of $\sigma_y(\tau)=1 \times 10^{-17}$, ie. a
flicker frequency noise at or below that level. The fractional frequency power
spectral density (PSD) is then
\begin{equation}\label{PSDspaceclock}
    S_{y_s}(f)= (2 \times 10^{-28} + 7.2\times 10^{-35}f^{-1})/ {\rm Hz}
\end{equation}
Though the flicker noise is likely to be lower than the projected accuracy, we
use this conservative estimate.

The ground clock stability is likely to be significantly better than the space
clock stability by the time the mission is launched. Best present stabilities
of optical frequency standards are already below $\sigma_y(\tau)=3\times
10^{-15} \tau^{-1/2}$ \cite{Hoyt06}, with accuracies at $3 \times 10^{-17}$
\cite{Fortier07}. Further rapid improvement of these numbers is expected. We
therefore estimate the ground clock noise by the time of mission operation at
$\sigma_y(\tau)=5\times 10^{-16} \tau^{-1/2}$ with a flicker component at $3
\times 10^{-18}$, so that the PSD is read
\begin{equation}\label{PSDgroundclock}
S_{y_g}(f)= (5 \times 10^{-31} + 6.5\times 10^{-36}f^{-1})/{\rm Hz}
\end{equation}

The noise on the spacecraft velocity of SAGAS is determined by the integrated 
noise of the on-board accelerometer. Although orbit modeling is likely
to improve on the raw accelerometer noise at low frequency, we use that as our 
conservative estimate for the purpose of this work. The accelerometer
noise specified in \cite{Wolf07} is 
$\sigma_a(\tau)=9\times 10^{-10} \tau^{-1/2}$~m/s$^2$ per axis for 3D measurements 
and $\sqrt{3}$ less when
measuring only along the direction of signal transmission, of interest here. 
An absolute accelerometer based on cold atom technology is used to avoid
long term drifts and biases (see \cite{Wolf07}, section 3.1 for details). 
The expected absolute accuracy is $5\times 10^{-12}$~m/s$^2$ taken again as
the upper limit of the flicker acceleration noise. 
This translates into a velocity PSD of
\begin{equation}
\label{PSDspaceV}
S_{v_\rs/c}= (1.5 \times 10^{-37}f^{-2} + 5.1\times
10^{-42}f^{-3})/{\rm Hz}
\end{equation}

For radio-frequency Doppler ranging, one of the dominant noise sources at low
frequency is the uncorrected motion of the 34~m DSN antenna and of the station
location itself \cite{Armstrong03}. For the optical link ground telescopes the
motion of the mirror is likely to cause less of a problem, however the site
movement plays a similar role as in the radio-frequency case. At high
frequencies the motion of the ground station can be corrected using gravity
measurements, with best presently achieved measurement noise levels of about
$4\times 10^{-18}$~m$^2$/s$^4$/Hz when using superconducting gravimeters
\cite{VanCamp05}. Alternatively, positioning using global navigation systems
(GNSS) and/or Satellite Laser Ranging (SLR) and/or Very Long Baseline
Interferometry (VLBI) achieve sub-cm uncertainties \cite{IGS}. Typically, GNSS
positioning shows flicker noise in position over a wide range of frequencies
\cite{Larson07,Williams04,Ray07}, with best results at present at about
$S_x(f)=1\times 10^{-6}f^{-1}$~m$^2$/Hz. The noise level shows some dependence
on the number of visible satellites, and therefore further improvement is
expected with upcoming additional GNSS systems \cite{GALILEO}. SLR and VLBI
show white positioning noise \cite{Ray07}, but at higher levels than GNSS at
the frequencies of interest here ($10^{-6}-10^{-5}$~Hz). Combining local
gravity measurements with GNSS positioning we obtain as the minimum noise on
ground velocity
\begin{eqnarray}\label{PSDgroundV}
\nonumber
&&S_{v_\rg/c} = 1.1\times 10^{-36}{f^{-2}\over{\rm Hz}}
\, ,\quad f>1.4\times 10^{-5} {\rm Hz}\\
&&S_{v_\rg/c} = 4.4\times 10^{-22}{f\over{\rm Hz}}
\, ,\quad f\leq
1.4\times 10^{-5} {\rm Hz}
\end{eqnarray}
In (\ref{PSDgroundV}) we have assumed that superconducting gravimeters display
white noise down to $10^{-5}$~Hz, whereas the spectra shown in \cite{VanCamp05}
only show white noise down to $10^{-3}$~Hz, with the measurements being
dominated by natural gravity fluctuations (the signal to be measured) at lower
frequencies. This has to be considered as a preliminary estimation of low
frequency noise unknown at present.

\begin{figure}[h]
\includegraphics[width=8 cm]{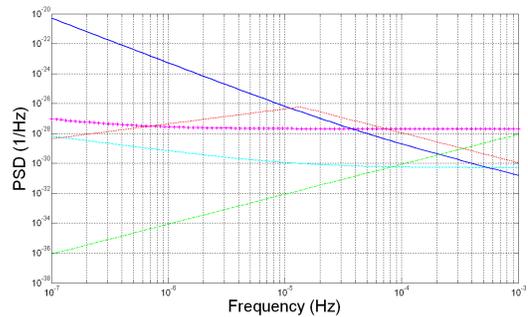}
\caption{Power spectral densities of the dominant noise sources on a one-way
link: $S_{y_s}$ (magenta-crosses), $S_{y_g}$ (light blue-dashed), $S_{v_\rs/c}$
(dark blue-solid), $S_{v_\rg/c}$ (red-dotted), $S_{y_\tropo}$
(green-dash-dotted).
\label{Figure4} }
\end{figure}

Typically tropospheric delay models at optical frequencies have millimetric
accuracy. Furthermore mapping functions that determine the variation of the
delay with elevation (of interest here, the constant part of the delay playing
no role on the frequency measurement) have been developed to sub-millimeter
accuracy and successfully tested on SLR data \cite{Mendes04}. Assuming that the
residuals from such models show white phase noise at the 1~mm level at 10~Hz
sampling (typical pulse rate of SLR and LLR stations) we obtain a frequency PSD
of
\begin{equation}\label{PSDtropo}
   S_{y_\tropo}(f)= 8.7 \times 10^{-23}f^{2}/{\rm Hz}
\end{equation}
Note, that we pessimistically ignore correlations at high frequencies, which,
given the slow motion of the satellite in the sky, should lead to decreased
high frequency noise.
Figure \ref{Figure4} summarizes the noise sources on a one way link discussed
above.

\subsection{Measurement noise on ranging and timing}

The noise on the ranging and timing observables defined in (\ref{timederiv}) is
a combination of the noise affecting the individual links, which are clearly
correlated when forming the ranging or timing observables. Taking into account
those correlations for the general case $t_2\neq t_3$ we obtain
\begin{eqnarray}
\label{PSDO+}
S_{y_\rr}(f) = \frac{1}{2}
&&\left\{ \left( 1 - \cos (2 \pi f T_{14}) \right) S_{y_\rg}(f) \right. \\
&&+ \left(1 - \cos (2 \pi f T_{23}) \right) S_{y_\rs}(f) \nonumber \\
&&+ \left(1 + \cos (2 \pi f T_{23}) \right) S_{v_\rs/c}(f) \nonumber \\
&&+ \left(1 + \cos (2 \pi f T_{14}) \right) S_{v_\rg/c}(f) \nonumber\\
&&+ \left. \left(1 + \cos (2 \pi f T_{14}) \right) S_{y_\tropo}(f) \right\} \nonumber \\
\label{PSDO-}
S_{y_\rt}(f) = \frac{1}{2}
&&\left\{ \left( 1 + \cos (2 \pi f T_{14}) \right) S_{y_\rg}(f) \right. \\
&&+ \left(1 + \cos (2 \pi f T_{23}) \right) S_{y_\rs}(f) \nonumber \\
&&+ \left(1 - \cos (2 \pi f T_{23}) \right) S_{v_\rs/c}(f) \nonumber \\
&&+ \left(1 - \cos (2 \pi f T_{14}) \right) S_{v_\rg/c}(f) \nonumber\\
&&+ \left. \left(1 - \cos (2 \pi f T_{14}) \right) S_{y_\tropo}(f) \right\} \nonumber
\end{eqnarray}
where we have defined $T_{ij}\equiv t_j-t_i$ (cf. fig. \ref{Figure1}).

Figures 5 show the resulting noise PSD of the timing and ranging observables
for the special cases $T_{23}=0$ (coincidence of up and down signals at the
satellite) and $T_{14}=0$ (coincidence of up and down signals on the ground).
More generally, (\ref{PSDO+}) and (\ref{PSDO-}) can be used to obtain the noise
spectra of $y_\rr$ and $y_\rt$ for arbitrary values of $T_{23}$ and $T_{14}$,
with the constraint $T_{14}-T_{23}=2 T$.

\begin{figure}[h]
\includegraphics[width=8 cm]{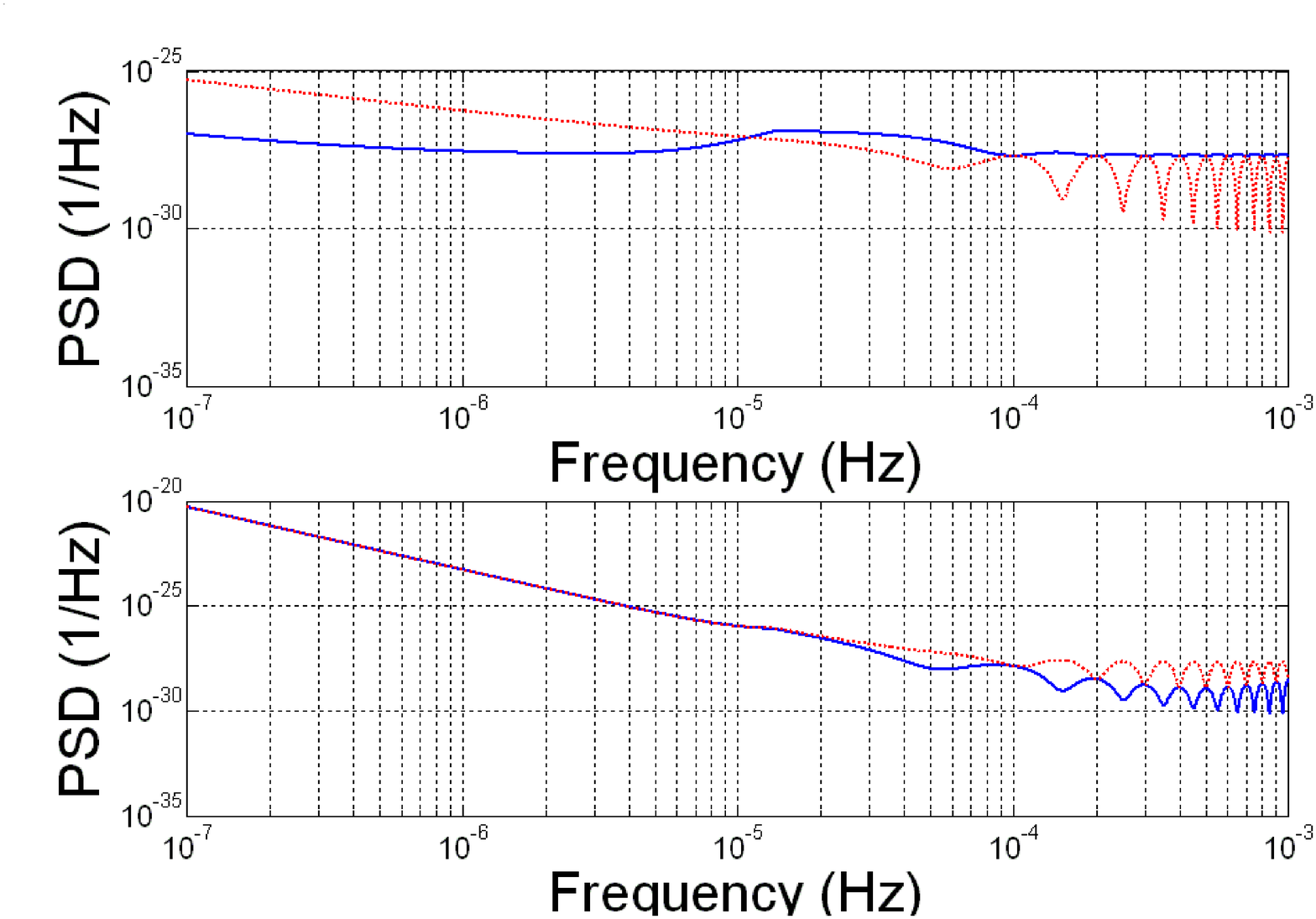}
\caption{PSD of the noise affecting the two observables $y_\rt$ (upper
graph) and $y_\rr$ (lower graph) for the special cases $T_{23}=0$ (blue-solid)
and $T_{14}=0$ (red-dotted), and for a spacecraft to ground distance of 10
AU.
\label{Figure5} }
\end{figure}

We note that the noise levels are significantly different, in particular at low
frequency, illustrating the potential gain one can expect from using the
optimal observable. In particular the observable $y_\rr$ with $T_{23}=0$ used
in "classical" Doppler ranging shows several orders of magnitude larger noise
at low frequency than the timing observable $y_\rt$ with $T_{23}=0$. This is
due to the cancelation of the onboard accelerometer noise in (\ref{PSDO-})
leaving only the onboard clock as the dominant noise contribution at low
frequency. This advantage has to be weighed against the different sensitivity
functions $b_\rr$ and $b_\rt$ as shown in fig. \ref{Figure3} in order to
determine the optimal observable as a function of Fourier frequency, which will
be the subject of the following section.

\section{Constraints on GW backgrounds}

The case of ranging or Doppler tracking ($y_\rr$) has been discussed in
numerous papers with the best bounds given by Doppler tracking of the Cassini
probe \cite{Armstrong06,Armstrong03}. One-way linking has also been studied
since it is involved in the extremely impressive bounds derived from pulsar
timing \cite{Kaspi94,Stairs03}. Here we will focus our attention on the case of
synchronization between remote clocks. As made clear by the discussion of the
preceding section, this points to experiments with excellent clocks at large
distances, and we will take the SAGAS project as an example. We begin by
discussing a somewhat simplified case illustrating the
advantages of the different observables and combinations, and then go on to
derive limits using the complete SAGAS noise sources as discussed in section 4.

In order to discuss the attainable performances, we use the spectra
associated with time derivatives of the phaseshifts induced by GW backgrounds
(eq.~\ref{sq}). These have to be compared to the phase variation induced by the
noise sources discussed in section 4. The equality of the two provides the
obtainable upper limit of $S_{\GW}$
\begin{equation}
S_{y}\left[ \omega \right] =\omega^{2}S_{\tau}\left[ \omega \right]
=\frac{5}{8}b\left[ \omega \right] S_{\GW}\left[ \omega \right] \label{defSy}
\end{equation}
with $S_y$ and $b$ having different forms for the different observables (cf.
equations (\ref{brtgeneral}), (\ref{PSDO+}), (\ref{PSDO-})).

For comparison with known bounds \cite{Abbott06}, we describe the
gravitational noise in terms of the reduced gravitational energy density
$\Omega_\GW $ commonly used to discuss the cosmic backgrounds 
($H_{0}\simeq 71 \ \mathrm{km} \mathrm{s}^{-1} \mathrm{Mpc}^{-1}
\simeq 2.3\times 10^{-18}~\mathrm{Hz}$ is the Hubble constant)
\begin{equation}
\Omega_\GW =\frac{10\pi ^2 f^3}{3H_0^2}S_{\GW}  \label{defOmega}
\end{equation}
Collecting (\ref{defSy}) and (\ref{defOmega}), we deduce the expression of the
bound obtainable on $\Omega_\GW $ from that of the sensitivity function $b$ and
the noise $S_y$
\begin{equation}
\Omega_\GW =\frac{16\pi ^2 f^3}{3H_0^2}\frac{S_y}{b} \label{boundOmega}
\end{equation}

\subsection{Illustration using a simplified case}

To illustrate how the obtainable limits can be optimised using the available
measurements and resulting observables, we first consider an idealized case
where only three noise sources play a significant role, the onboard clock and
accelerometer, and the ground clock. Furthermore, we will assume that all three
noises consist of only white noise, at the levels indicated by equations
(\ref{PSDspaceclock}, \ref{PSDspaceV} and \ref{PSDgroundclock}), ie. we will
only consider the first terms of those equations in the expressions
(\ref{PSDO+}, \ref{PSDO-}) for $S_y$, with all other terms set to zero.

Figure 6 shows the resulting limits on $\Omega_\GW$ as a function of frequency
for two satellite to ground distances: 6~AU, the distance of Cassini when the
GW experiment was carried out \cite{Armstrong03}, and 53~AU, the maximum
distance envisaged for the SAGAS mission \cite{Wolf07}. Limits are shown for
three observables: $\Omega_\rr$ is obtained using the "classical" Doppler
ranging observable as defined in (\ref{range}) with $t_2= t_3$. We recall that
this is the only observable available on space-probes equipped only with a
transponder (the case of all probes flown so far). $\Omega_\rt$ is obtained
using the timing observable defined in (\ref{timing}) again with the condition
$t_2= t_3$. $\Omega_\opt $ is calculated by adjusting $T_{23}$ in
(\ref{brtgeneral}, \ref{PSDO+} and \ref{PSDO-}) for each frequency in order to
minimize the obtained limit on $\Omega_\GW$.

In doing so, one can use either the ranging or timing combination, the obtained
optimal limits being identical (albeit for different values of $T_{23}$). That
property is the result of the periodic dependence of equations
(\ref{brtgeneral}, \ref{PSDO+} and \ref{PSDO-}) on $T_{23}$, which means that
at any given frequency one can find two values of $T_{23}$ for which the
ranging and timing combinations yield the same limit on $\Omega_\GW$ in
equation (\ref{boundOmega}). However, the assumption that the up and down
travel times are similar (see section 3.1) limits the allowed range of
$T_{23}$. Taking into account the maximum relative probe-earth velocity
($\approx 50$~km/s) we limit $T_{23}$ in the calculation of $\Omega_\opt $ so
that the up and down travel times do not differ by more than 1\%. We then
choose as $\Omega_\opt $ the lower of the two limits obtained from $y_\rr $ and
$y_\rt $ with a free choice of $T_{23}$ within the 1~\% limit mentioned above.

\begin{figure}[h]
\includegraphics[width=8 cm]{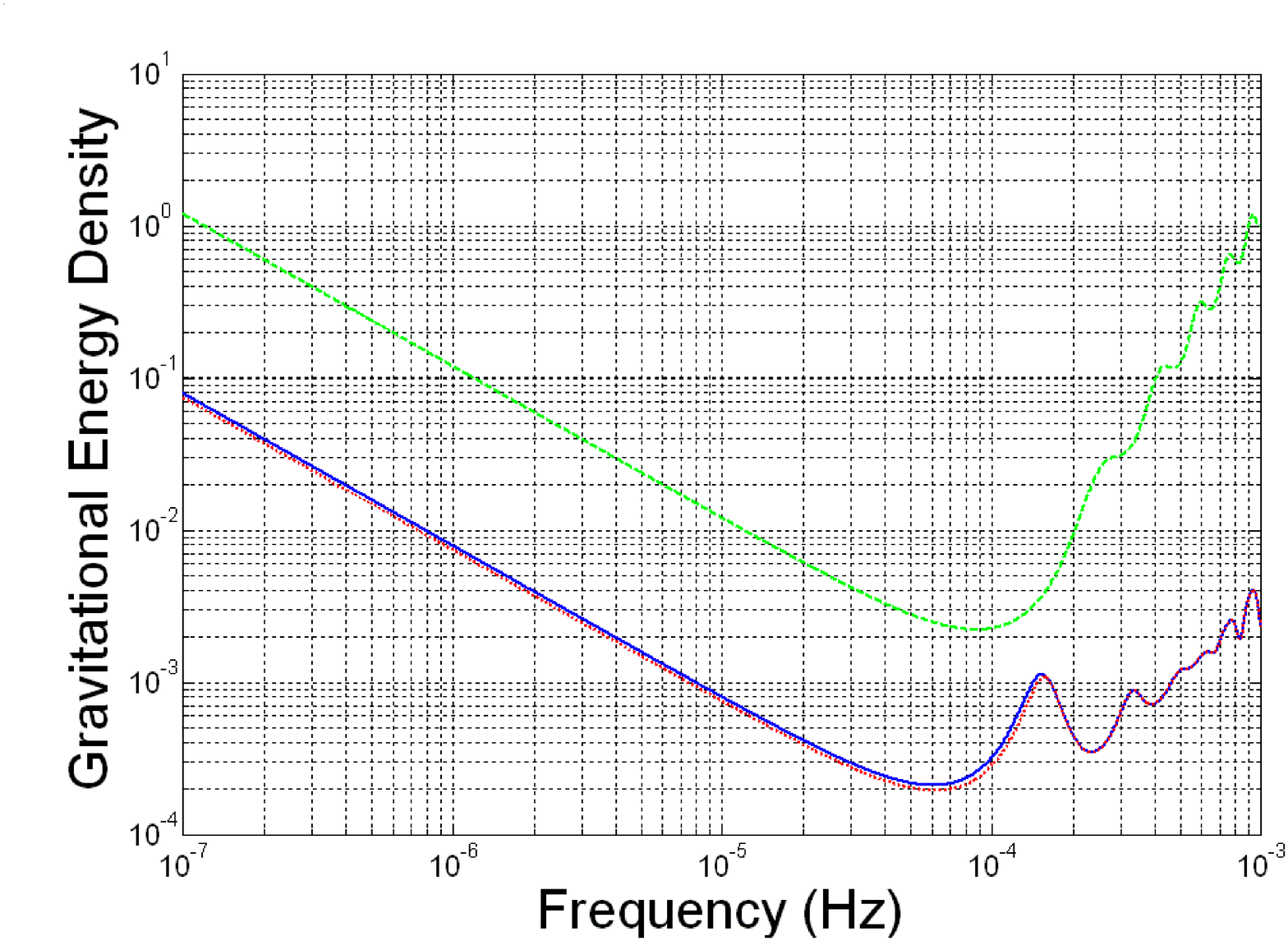}
\includegraphics[width=8 cm]{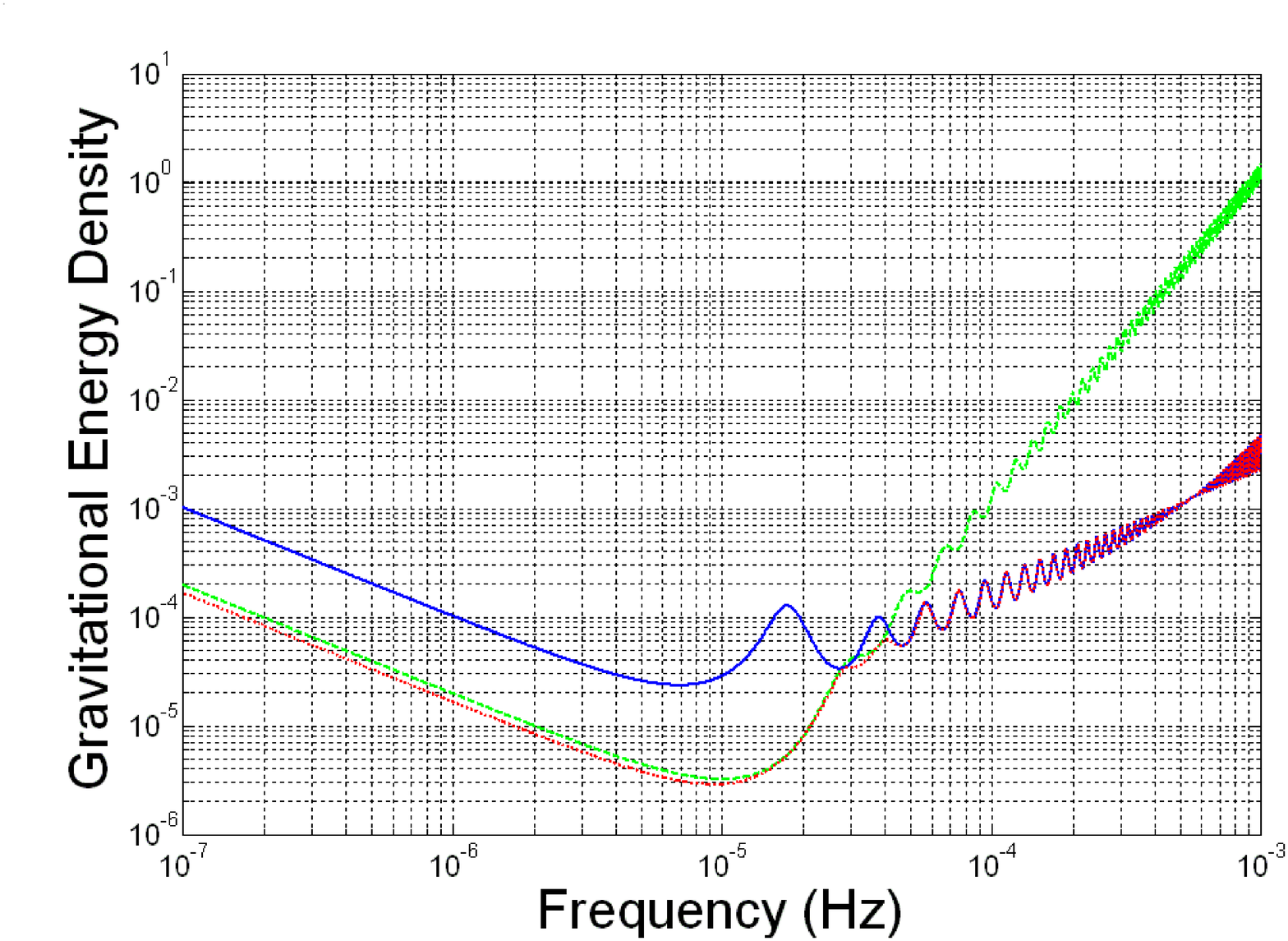}
\caption{$\Omega_\GW$ versus $f$, for a probe to Earth distance of 6~AU
(left) and 53~AU (right). The three curves show the limits on $\Omega_\rr $
(blue-solid), $\Omega_\rt $ (green-dashed) and $\Omega_\opt $
(red-dotted).
\label{Figure6} }
\end{figure}

The graphs on figure \ref{Figure6} can be understood qualitatively and
quantitatively when considering the expressions for the sensitivity functions
(\ref{brtgeneral}) and the overall noise of the observables (\ref{PSDO+}) and
(\ref{PSDO-}) that enter into the calculation of $\Omega_\GW$ in
(\ref{boundOmega}). We first discuss $\Omega_\rr $ and $\Omega_\rt $ in the
case $T_{23}=0$, and then come to $\Omega_\opt $.

At the low frequency limit ($\omega T << 1$), $b_\rr \propto (\omega T)^2$ and
$b_\rt \propto (\omega T)^4$ (see section 3.2). The noise at low frequency is
dominated by the space probe motion for $S_{y_\rr }$ and is thus proportional
to $\omega^{-2}$, but that contribution is entirely canceled in $S_{y_\rt }$
because of the condition $T_{23}=0$ in (\ref{PSDO-}), so $S_{y_\rt }\propto
\omega^0$ (clock noise only). This leaves an overall $\omega^{-1}$ dependence
for both $\Omega_\rt $ and $\Omega_\rr $, clearly displayed in both plots of
figure \ref{Figure6} at low frequency. Also, $\Omega_\rr $ is significantly
lower than $\Omega_\rt $ at 6~AU, while the inverse is true at 53~AU (at low
frequency). This is caused by a tradeoff between the difference in sensitivity
functions and the involved noise sources. We find more specifically and to
leading order
\begin{equation}
\lim_{\omega T \ll 1}\frac{\Omega _\rr }{\Omega _\rt }= \frac{1}{4}
\left(\omega T\right)^2 \frac{S_{v_S}}{S_{y_S}}
\end{equation}
where $S_{v_S}$ and $S_{y_S}$ are the first terms of (\ref{PSDspaceV}) and
(\ref{PSDspaceclock}) respectively, ie. $S_{v_S}/S_{y_S}\simeq 3\times 10^{-8}
\omega^{-2}$. Thus for large distances (greater than $\simeq 20$~AU with SAGAS
figures) the low frequency asymptote is lower for timing than for ranging,
leading to the observed inversion of $\Omega_\rr $ and $\Omega_\rt $ when
passing from 6~AU to 53~AU.

At the other end of the spectrum ($\omega T >> 1$), ranging outperforms timing
(ie. $\Omega_\rr  < \Omega_\rt )$ at both distances. This can be easily
understood when considering only the involved noise sources, as the sensitivity
functions show oscillatory behavior and differ at most by a factor~3 (see
section 3.2). At high frequency (between $10^{-4}$~Hz and $10^{-3}$~Hz),
$S_{y_\rt }$ is dominated by the space clock, but that contribution is entirely
canceled in $S_{y_\rr }$ because of the condition $T_{23}=0$ in (\ref{PSDO+}),
leaving only a combination of space probe motion and ground clock noise. As at
high frequency the space clock noise is significantly higher than that of the
space probe motion or the ground clock (see figure \ref{Figure4}) this leads to
the observed advantage of $\Omega_\rr $ over $\Omega_\rt $. Note also, the
difference in slope between $\Omega_\rr $ and $\Omega_\rt $ at high frequency,
particularly visible at 53~AU, which can be easily understood from the slopes
of the spectra of the different noise contributions ($S_{y_g}\propto f^0$,
$S_{v_S/c}\propto f^{-2}$).

The lowest limits are obtained in the intermediate region ($\omega T \simeq 1$)
with very different results for the two distances. As expected, even when using
only the "classical" ranging observable ($\Omega_\rr $ on figure
\ref{Figure6}), limits improve with distance by about the ratio of distances
(about an order of magnitude in the present case). However, it is clearly seen
that almost another order of magnitude can be gained when taking advantage of
the timing observable ($\Omega_\rt $ in figure \ref{Figure6}), only available
when using space probes equipped with an onboard clock and a two-way
electromagnetic link.

In that case, one can not only choose between $y_\rr $ and $y_\rt $, but also
adjust the value of $T_{23}$ in order to optimize the measurement for any given
frequency. The result of such an optimization, $\Omega_\opt $, is shown in
figure \ref{Figure6}. As expected it is below $\Omega_\rr $ and $\Omega_\rt $
at all frequencies and for both distances. Although the overall improvement is
not spectacular, one obtains the "best of both worlds", in particular at 53~AU
where $\Omega_\opt $ follows $\Omega_\rt $ at low frequency and $\Omega_\rr $
at high frequency. We notice a slight improvement on $\Omega_\rt $ at low
frequency and 53~AU, which can be understood by considering the series
expansion of
 the sensitivity function $b_\rt $ in (\ref{brtgeneral}).
Additionally to the term in $(\omega T)^4$ present in the case $T_{23}=0$ (see
section 3.2) one now obtains a term proportional to $(\omega T )^2(\omega
T_{23})^2$ which can be significantly larger. However, when $T_{23}\neq 0$, low
frequency noise from the space probe motion is added, the trade off between the
two leading to the small improvement of $\Omega_\opt $ over $\Omega_\rt $ seen
on figure \ref{Figure6}.

In conclusion, the simplified case used in this section illustrates the
advantages of having an onboard clock and a two-way link, which allows
one to "fine tune" the data analysis as a function of the expected signal and
the noise sources affecting the raw measurements. In this example, the
sensitivity to GW backgrounds at large distance (53~AU) is improved by about an
order of magnitude over the "classical" case $\Omega_\rr $ by choosing the
optimal combination of the available measurements on ground and onboard the
space probe. Similar (up to a factor 20) improvements are observed when taking
into account all noise sources discussed in section~4.

\subsection{SAGAS limits on GW backgrounds}

\begin{figure}[h]
\includegraphics[width=8 cm]{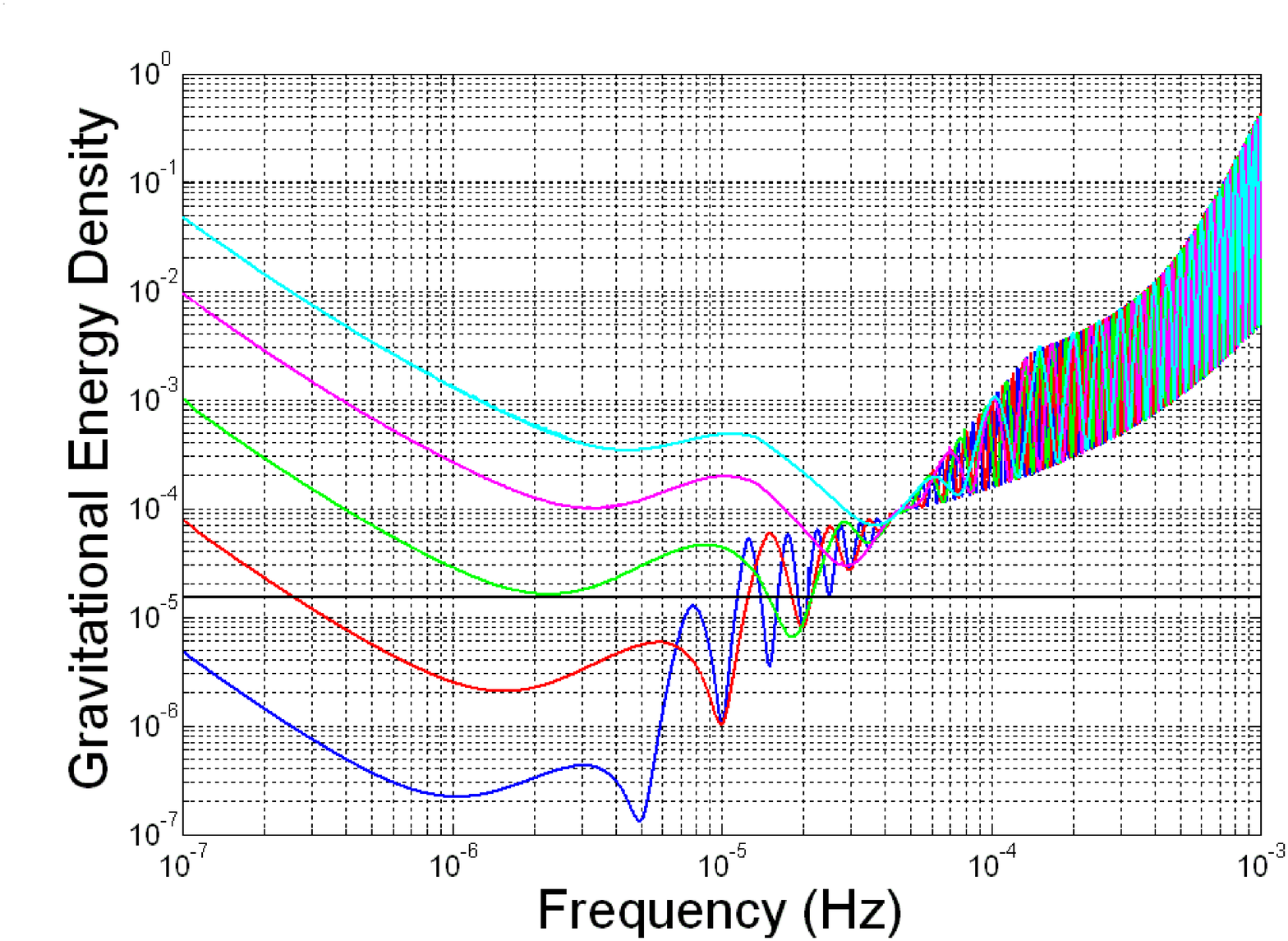}
\caption{$\Omega_\opt $ versus $f$, for probe to Earth distances (top to
bottom) of 20~AU (light blue), 30~AU (magenta), 53~AU (green), 100~AU (red) and
200~AU (dark blue). The horizontal black line indicates the BBN
bound.
\label{Figure7} }
\end{figure}

We now repeat the calculations described in the previous section, using the
example of SAGAS including all noise sources described in section 4. For
clarity, we show only the resulting optimal limits $\Omega_\opt $ for a range
of distances (see figure~\ref{Figure7}). The Big Bang Nucleosynthesis (BBN)
bound, which corresponds to a flat floor with $\Omega _\BBN\sim 1.5\times
10^{-5}$ has been drawn for comparison.

As expected, the sensitivity to GW is improved by going to large distances. The
frequency at which the lowest limits are reached is typically in the region
where $fT\sim 1$ ie. decreasing with increasing distance. For SAGAS (projected
distance $\sim 53$~AU) the lowest obtained limit is about $\Omega_{\GW}\leq
7\times 10^{-6}$ around $2\times 10^{-5}$~Hz and essentially determined from
the timing observable. It is more than 3 orders of magnitude below the best
directly measured limits in the $10^{-6}$ to $10^{-4}$ Hz band, obtained from
the Cassini probe \cite{Armstrong03}, and about a factor 2 below the BBN bound.
At larger distances the timing measurement could even approach the pulsar bound
around a few $10^{-8}$, but at significantly higher frequencies (the pulsar
bounds are at a few nHz \cite{Kaspi94}).

\subsection{Discussion}

Existing bounds on stochastic GW backgrounds are spread over a huge frequency
range from $10^{-18}$ Hz to $10^3$ Hz, corresponding to cosmological bounds
obtained from measurements of the 3 K microwave background (COBE) at the lower
end, and modern ground based GW detectors (LIGO, VIRGO) at the upper end (see
e.g. figure 14 of \cite{Abbott06}). This large frequency range is patchily
covered, with COBE limits at $10^{-18} - 10^{-16}$ Hz, pulsar bounds around
$10^{-9} -10^{-8}$ Hz, spacecraft Doppler ranging covering three orders of
magnitude ($10^{-6} - 10^{-3}$ Hz), and GW detectors setting limits around
$10^{2}$ Hz. This is complemented by an indirect upper limit derived from
models of BBN which corresponds to a flat floor of $\Omega_{\GW} \leq 1.5
\times 10^{-5}$ at all frequencies $\geq 10^{-10}$ Hz. The latter is already
outperformed by the pulsar limits at low frequency and is expected to be
outperformed at high frequency by LIGO and VIRGO measurements in the near
future.

In this landscape, limits obtained from spacecraft tracking play an important
role as they fill a large part of the gap in frequency between the pulsar
limits and those obtained from ground based detectors. Unfortunately, the
obtained bounds are presently limited to $\Omega_{\GW} \leq 0.025$
\cite{Armstrong03}. Improvements in this band will be particularly useful,
especially when they will approach or surpass the BBN bound, as would be the
case with future missions like SAGAS. As shown above, we expect that such
missions could provide limits on $\Omega_\GW$ down to parts in $10^{-6}$ for
SAGAS and below for missions at even larger distances.

More generally, it is important to obtain experimental constraints on
$\Omega_\GW$ at all frequencies, as many of the models that predict such GW
backgrounds are frequency dependent (eg. cosmic strings models, pre Big Bang
models,...) and only poorly constrained in the presence of frequency gaps. In
that respect the future space interferometric GW detector (LISA) plays an
important role, as it should provide extremely low limits (down to
$\Omega_{\GW} \leq 10^{-13}$) in the still largely unconstrained frequency
range of $10^{-4}$ Hz to $10^{-1}$ Hz. It should even be able to observe the
astrophysical GW background from an ensemble of galactic binary stars,
estimated to be too low for any other present or planned detector, but within
the reach of LISA.

\section{Conclusion}

Doppler ranging to distant space probes provides the presently most stringent
upper bounds on GW between $10^{-6}$ and $10^{-3}$ Hz. Those bounds are
obtained by "passive" ranging, where the space probe only serves as a
"reflector" of the signal emitted from the ground. We have shown that the
sensitivity can be significantly improved when having a clock onboard,
so that the up and down signals are independent (asynchronous link) and can be
combined in an optimal manner adapted to the signal to be measured and the
noise affecting the link. We have derived explicit expressions for the
sensitivity of all possible link combinations to a GW background. Using the
example of the SAGAS project, we have evaluated the sensitivity of such a
mission to GW backgrounds for optimal signal combinations and as a function of
distance, with a potential improvement by over three orders of magnitude on
best present limits.

Let us notice the similarities between the calculations of the present paper
and those previously devoted to the effect of stochastic GW backgrounds on
inertial sensors built on atomic interferometry
\cite{Reynaud07,Lamine02_06}. The sensors of interest in the
present paper are the atomic clocks the indications of which are compared
through electromagnetic links. As these links cannot be protected against the
action of GW backgrounds, there exists an ultimate noise in clock
synchronisation due to the presence of this universal fluctuating environment.
It has been shown in the present paper that timing can be more sensitive to
this environment than ranging, provided that extremely large distances are
considered, as it is the case in the SAGAS project.

In our estimations we have chosen a conservative approach where the noise on
the spacecraft motion is determined solely by the measurement noise of the
onboard accelerometer. Previous deep space probes, in particular the Cassini
mission \cite{Armstrong03}, did not have an accelerometer on board, and
all non-gravitational accelerations acting on the probe where determined by
fitting acceleration models to the ranging data. The PSD of the residuals is
most likely dominated by ground station and antenna motion at low frequency,
and in particular around the diurnal frequency and its harmonics (see figure~1
of \cite{Armstrong03}). For a mission like SAGAS this suggests an analysis
strategy based on the cancellation of the ground station motion rather than
that of the space probe. In frequency regions where ground station noise is
dominant one would use the timing observable (\ref{timing}) giving rise to
$S_{y_\rt }(f)$ of (\ref{PSDO-}), but with the condition $T_{14}\simeq 0$
(coincidence of up and down signals at the ground antenna). As can be easily
seen from (\ref{PSDO-}), this leads to cancellation of noise from the ground
station motion and the troposphere, leaving space clock instabilities and space
probe motion as the dominant noise sources (see figure \ref{Figure4}). To
evaluate the limits obtained in this scenario requires a more detailed
investigation of the space probe motion, the effect of fitting acceleration
models, the improvements in the fits from in situ acceleration measurements,
etc..., which are beyond the scope of this paper. Nonetheless, this alternative
approach well illustrates the versatility of using an asynchronous link
that allows choosing the optimal data combination strategy, even after launch,
and as a function of the observed noise levels.

Finally, we point out that this data combination strategy can be adapted and
optimized for any signal that is to be measured. The GW backgrounds discussed
in this paper give rise to sensitivity functions (\ref{brtgeneral}) which enter
the parameter $\Omega_\GW$ in (\ref{boundOmega}) together with the noise
(\ref{PSDO+}), (\ref{PSDO-}). That parameter is then optimized over a broad
frequency range by varying $T_{23}$ or $T_{14}$. A similar procedure can be
used for other science objectives (e.g. test of the gravitational time delay
during occultation, measurements of planetary gravity, trajectory determination
during fly by, etc...) by deriving the appropriate sensitivity functions and
calculating the parameters to be optimized. It is likely to allow significant
improvements in those measurements as well.

\begin{acknowledgments}
Helpful discussions with Jim Ray of NGS/NOAA concerning GNSS,
SLR and VLBI positioning noise are gratefully acknowledged. Loic Duchayne
acknowledges financial support from ESA, EADS-Astrium and CNES.
\end{acknowledgments}

\end{document}